\documentclass[utf8,latin1,onecolumn,aps,pra,preprint]{revtex4-1}
\usepackage{amsmath,amssymb,bm,graphicx,epsfig,psfrag,ulem,color,siunitx,float,upgreek,ae,aecompl}
\usepackage[utf8]{inputenc}
\graphicspath{{Figs/}}

\begin{document}

\title{Shape and size measurements of nonequilibrium Bose-Einstein condensates using image processing}

\author{J.~P.~G.~Venassi, V.~S.~Bagnato, G.~D.~Telles}
\affiliation{Instituto de F\'{\i}sica de S\~{a}o Carlos,
Universidade de S\~{a}o Paulo, Av. Trabalhador S\~{a}o-carlense, 400, Pq. Arnold Schimidt, 13566-590, S\~{a}o Carlos, SP, Brazil}

%\markboth{Journal of \LaTeX\ Class Files,~Vol.~18, No.~9, September~2020}%
%{How to Use the IEEEtran \LaTeX \ Templates}

\begin{abstract}
Bose-Einstein condensates have been the subject of intense research in recent years due to their potential applications in quantum computing and many other areas. However, measuring the shape and size of out-of-equilibrium Bose-Einstein condensates is a challenging task that requires sophisticated image processing techniques. We propose to study perturbed BEC based on general concepts of analysis, which are widely used in the image processing community. The mathematical basis underlying the algorithms is quite general and independent of the type of image studied. The morphological changes observed in the perturbed atomic clouds as a result of excitation amplitude were observed in a consistent manner. And the spatial expansion of the atomic clouds under free fall shows some symmetry, but it was only observed under certain conditions
\end{abstract}

%\begin{IEEEkeywords}
%Class, IEEEtran, \LaTeX, paper, style, template, typesetting.
%\end{IEEEkeywords}
\maketitle
%\nocite{*}
\section{Introduction}
Bose-Einstein condensates (BECs) are a unique state of matter created by cooling a gas of neutral atoms to extremely low temperatures. Bose-Einstein condensates have been the subject of intense research in recent years due to their potential applications in quantum computing, to simulate complex quantum systems, as well as to develop ultra-precise sensors~\cite{Ketterle99,Bagnato15bec,Ketterle99}. BECs are highly sensitive to external perturbations, which may cause them to radically change their shape and size~\cite{Bagnato2013,Vivanco23}. As a result, it can be difficult to accurately determine these properties. In fact, investigating most features of out-from-equilibrium Bose-Einstein condensates is challenging, and it may require more elaborated digital image processing techniques than those already in use. 

The accurate determination of the shape and size of out-of-equilibrium Bose-Einstein condensates (BECs) is a challenging task for a number of reasons.
First, out-of-equilibrium BECs are often very transient, meaning that they only exist for a short period of time. This makes it difficult to image them accurately.
Second, they can be very sensitive to external perturbations, such as stray light or magnetic fields. This can make it difficult to isolate them from the environment and to obtain accurate measurements~\cite{Bagnato2013,Vivanco23}
Third, the shape and size of a perturbed BEC can be very complex and can change rapidly. This makes it difficult to track the evolution of the BEC over time.
Nonetheless, there are a few techniques that can be used to measure the shape and size of out-of-equilibrium BECs. The standard technique is the time-of-flight (TOF) imaging, by which the BEC is first allowed to expand, falling under gravity, for a short period of time and then imaged using a resonant probe laser beam~\cite{Ketterle99,Szczepkowski2009}. The shape and size of the BEC can then be determined from the images.
Another common technique that can be used to measure the shape and size of out-of-equilibrium BECs is to use phase-contrast imaging. In this technique, the BEC is illuminated with a laser beam detuned from the resonant frequency of the atoms, and the resulting interference pattern can be used to image the BEC~\cite{Ketterle99}.
The choice of technique for measuring the shape and size of a out-of-equilibrium BEC will depend on the specific experimental conditions. However, all the techniques mentioned above have their own advantages and disadvantages.
Here are some additional challenges that may be encountered when measuring the shape and size of out-of-equilibrium BECs:
The presence of  thermal atoms change the dynamics of a condensate in such a way that both radial and axial atomic cloud widths get smaller compared to expansion of pure BECs~\cite{Reinaudi07,Szczepkowski2009,Gajdacz2013,Hueck17}. This causes the condensate's aspect ratio to vary, becoming larger for a smaller condensate fractions (or larger thermal clouds). And, the finite resolution of the imaging system can introduce errors in the measurements. The kinematics and dynamics of the BEC can be affected by the measurement process itself. Despite these challenges, measuring the shape and size of out-of-equilibrium BECs is an important task that can provide valuable insights into some hidden properties.

\section{Experimental setup and data acquisition}
All the images used in this study comprise atomic $^{87}\rm{Rb}$ BECs produced in the $\left|F=2,m_{F}=+2\right>$ hyperfine state, with a small thermal atom fraction in a cigar-shaped QUIC magnetic trap, and the detailed description of the full sequence can be found elsewhere~\cite{Vivanco23}. The amplitude unit is peak-to-peak volts ($V_{pp}$) detected across a reference resistor placed in series with the excitation coils, and $1V_{pp}\approx$~47\unit{\micro\tesla} superposed to the QUIC trap for the present experiment. In previous experiments we used different excitation coils system, which had different calibration~\cite{Bagnato2013}. The time-varying extra gradient field oscillates with a frequency of about 190~Hz during an excitation time $t_{\rm exc}$, slightly deforming the static magnetic potential shape, which perturbs the trapped atomic cloud. After $t_{\rm exc}$ the system is then left to evolve in-trap for a time $t_{\rm hold}$. This procedure is used to generate the nonequilibrium BECs, where the kinetic energy rules over the inter-particle interaction energy.

In out-of-equilibrium regimes, the perturbed atom density may not vary smoothly over space and the optical resolution of an \textit{in-situ} imaging microscope might not be ideal to accurately reveal the real space density (high-frequency) modulations caused by the external perturbation. On the other hand, an expanded atom density will magnify the density modulations, allowing for a more detailed analysis, though its Fourier transform do not represent the linear momentum distribution of the (previously) trapped particles, since the “magnified” atom density does not map to the \textit{in-situ} density in a simple manner.
\begin{figure}[htp]
    \centering
    \includegraphics[width=7cm]{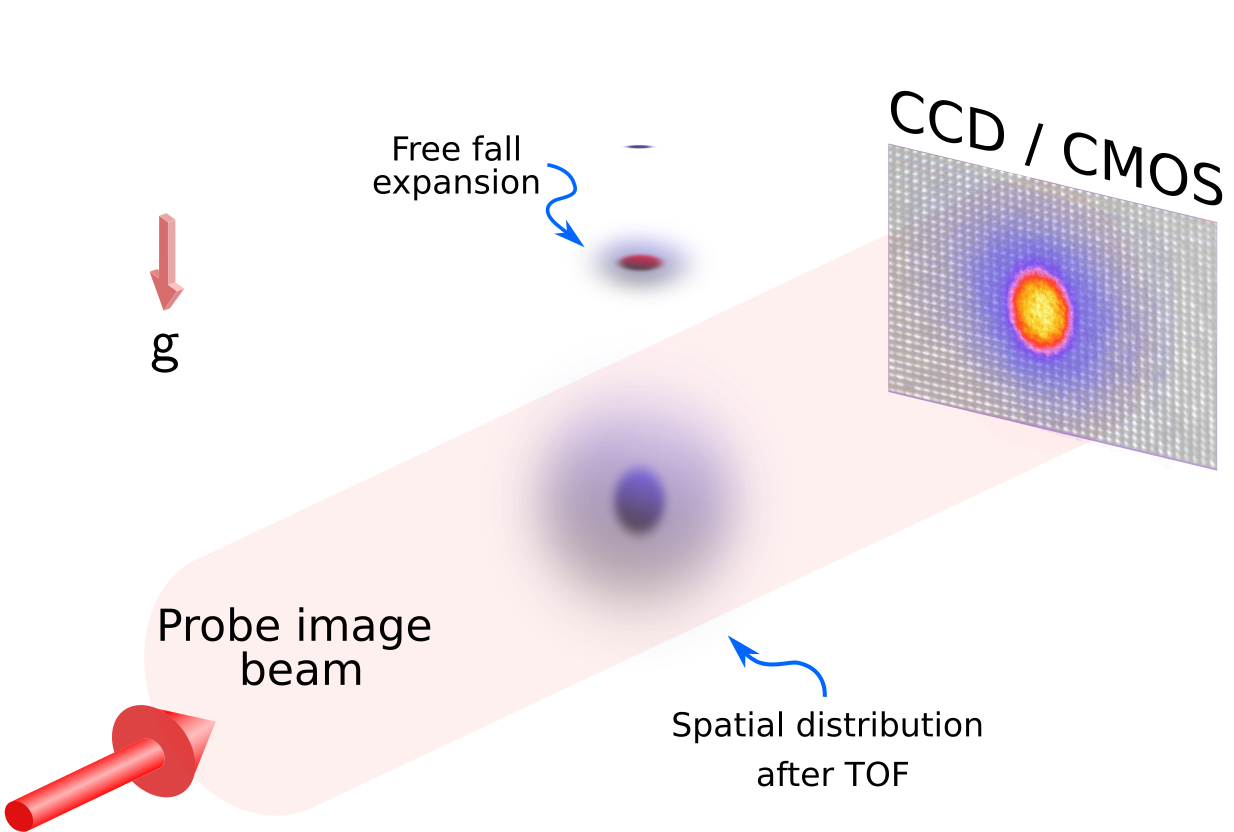}
    \caption{Diagram showing the basic idea of the time-of-flight imaging technique used to acquire the images used in this work.}
    \label{fig:tofimg}
\end{figure}
\section{Image processing techniques}
The purpose of imaging and image processing is to provide reliable shadowgraphs with the density distributions of the atomic clouds, either trapped or in ballistic expansion as those used in this work. All properties of condensates and thermal clouds are inferred from these density distributions. This is done by comparing the measured distributions with the results of models of the atomic gas. All imaging techniques measure the column density of the cloud along the probe laser propagation direction. Using destructive imaging such as TOF to probe the atomic gas implies that, for each data point in the accumulated time-series, the experiment is cycled through the entire loading and cooling cycle. Thus, such data time-series are susceptible to shot-to-shot fluctuations. 
In summary, the methods are able to produce good quality images to be analyzed, and brings accurate information about the shape and size of the atomic distributions, but they are still subject to limitations such as weak signal-to-noise strength and other sources of error, such as the lack of theoretical framework to fit the spatial atomic distribution.
The interpretation of time-of-flight images to determine temperature and chemical potential assumes a sudden, ballistic, and free expansion (see Fig.\ref{fig:tofimg}). A variety of effects may invalidate this assumption. These include residual magnetic fields, such as a slow and/or unbalanced trap shut-off during the free-expansion period. Another constraint is that thermal clouds released from anisotropic traps can expand anisotropically due to collisions soon after the trap is switched off. Also, in TOF starting from a cigar-shaped potential, the thermal cloud expands isotropically once $t_{TOF}>\omega_z^{-1}$, while the condensate expands axially beyond its initial length only when $t_{TOF}>\omega_r/\omega_z^2$. For times $1<\omega_z t_{TOF}<\omega_r/\omega_z$, if the imaging is taken along a radial direction, the condensate optical density drops linearly with expansion time, while the non-condensate optical density drops quadratically.
There are also differences in measuring dynamic properties. One goal is measuring oscillations at the limit of zero-amplitude. It has been shown that the relative amplitude of shape oscillations observed in TOF was much greater than what would be observed \textit{in-situ}~\cite{Dalfovo99,Ketterle99}. Thus, TOF imaging allows for the observation of smaller amplitude oscillations. On the other hand, this benefit of TOF imaging may be offset by slower data acquisition and greater susceptibility to noise.
On the other hand, there are a number of standard, well developed and understood, standard image processing techniques commonly used in computer vision that may be used to accurately measure the shape and size of Bose-Einstein condensates (BECs). Some of the most common techniques include~\cite{gonzalez08}:
\begin{itemize}
    \item \textit{Thresholding}: This technique involves setting a threshold value for the intensity of the image. All pixels with an intensity above the threshold value are considered to be part of the BEC, while all pixels with an intensity below the threshold value are considered to be background.
    \item \textit{Blob detection}: This technique involves identifying connected regions of pixels that have a similar intensity. The BEC can then be identified as the largest blob in the image.
    \item \textit{Morphological operations}: This technique involves performing operations on the image, such as dilation, erosion, and opening, to remove noise and improve the contrast of the image. This can make it easier to identify the BEC in the image.
    \item \textit{Moments}: This technique involves calculating the moments of the image, which are a set of statistics that describe the distribution of the intensity in the image. The shape and size of the BEC can then be determined from the moments.
    \item \textit{Hu moments}: This technique is a special case of moments that is specifically designed for shape analysis. The Hu moments are invariant to translation, rotation, and scaling, which makes them well-suited for measuring the shape of the BEC.
   
\end{itemize}
Here are some additional considerations when using image processing techniques to measure the shape and size of BECs:
The choice of threshold value is important, as it can affect the accuracy of the measurements.
The blob detection algorithm should be chosen carefully, as different algorithms can produce different results.
The moments should be calculated using a robust method, as the shape of the BEC can be affected by noise in the image.
The morphological operations should be used carefully, as they can distort the shape of the BEC.
Despite these considerations, image processing techniques are a powerful tool that can be used to accurately measure the shape and size of BECs.
\section{Results and Applications}
\begin{figure}
    \centering
    \includegraphics[width=1.0\linewidth]{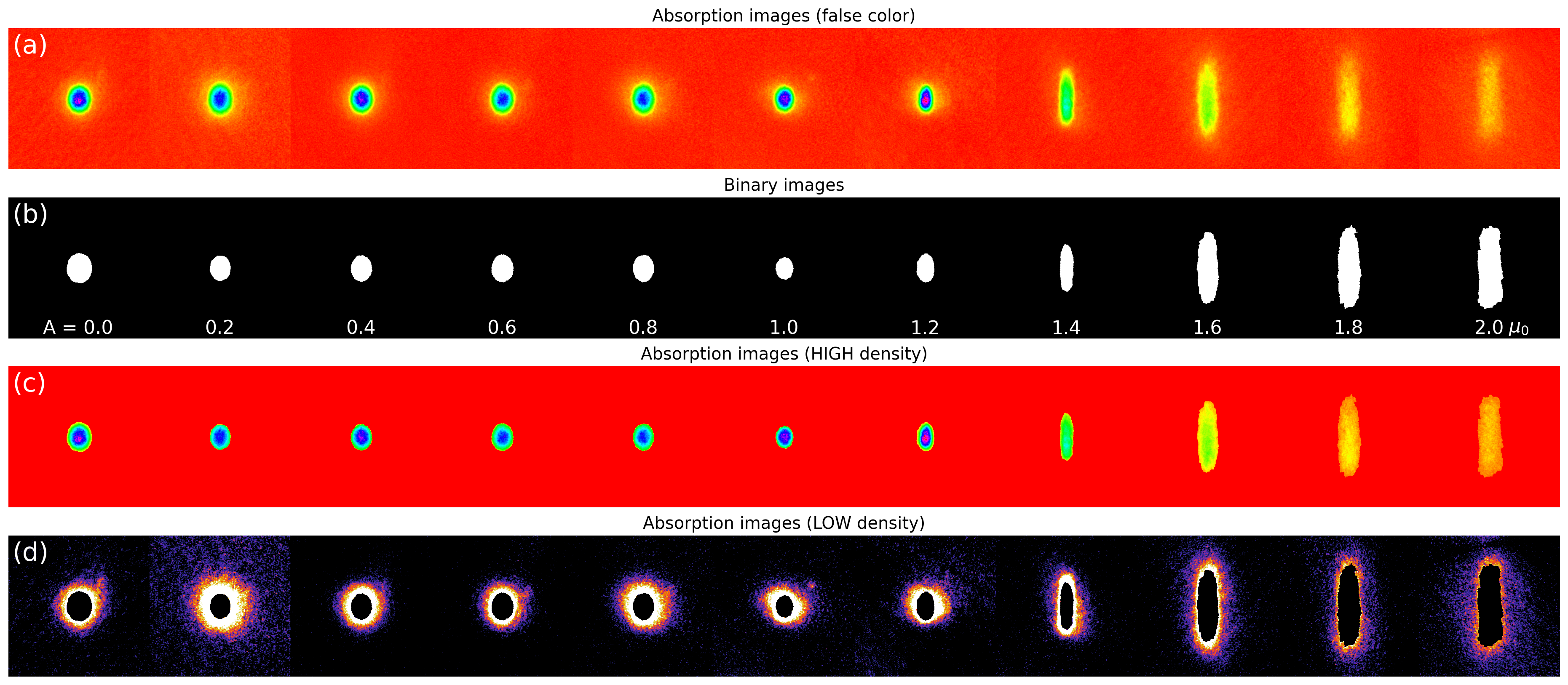}
    \caption{This image panel presents the whole preprocessed sequence of ROI images of the atomic clouds, from top to bottom. We start by presenting the cropped, normalized, false color experimental images acquired after 24~ms TOF, as depicted in Fig.\ref{fig:tofimg}. The binary images are shown in the $2^{nd}$ line (b) and were used to create masks which were applied over the images in the $1^{st}$ line (a) to generate the high (c) and low (d) optical density shadow graphs, shown in false colors. These last two lines in the panel are meant to show more clearly the remaining spatial atomic distributions spread inside/outside the binary images borders.}
    \label{fig:ImgPans}
\end{figure}

Recently, we have been running experiments where a rubidium Bose-Einstein condensate is magnetically perturbed, and pushed out of equilibrium. This process may drive the BEC towards turbulent states, which has been studied in our laboratory. Turbulent BECs present large density fluctuations, which can be observed by imaging the atomic density distribution, probed by optical absorption during expansion under TOF.
We will now present the main ideas followed to process and analyze the image data acquired during the experiments. The first step was to analyze the full normalized image to determine the field-of-view (FOV) available for the processing. We  begin by showing the initial part or preprocessing where the image files were simply scanned to locate the atomic clouds' center-of-mass (COM) or simply centroid, and then crop the image around it determining the region-of-interest (ROI) square box. This preprocessing is a fully automated algorithm written in Python. In this step, the challenging task accomplished was to properly normalize the image, assuming light saturation conditions~\cite{Reinaudi07,Hueck17}, and then to automatically determine the right gray level threshold needed to create a binary version~\cite{SciktImg14} of the (cropped) normalized images. Most of the time was spent on fine tuning the algorithm to work flawlessly and we have used the dataset acquired after 24~\unit{ms} TOF and 30~\unit{ms} of hold time. Then there was another step needed to optimize it to work well over the whole dataset, ranging from 24-30~ms TOF (??).
In Fig.\ref{fig:ImgPans} we present a sequence of normalized absorption images of a $^{87}$Rb BEC driven out of equilibrium, showing how the different amplitudes change the spatial atomic distribution as the perturbing magnetic field amplitude grows from the left to the right. The images presented in Fig.\ref{fig:ImgPans} were all acquired in free fall, after 24 ms TOF.% On the top montage, we show the full picture, presenting the center-of-mass motion and morphing, and on the bottom panel we zoom in to show the region of interest (ROI) of 925\unit{\square\um}.} 
\begin{figure}
    \centering
    \includegraphics[width=0.8\linewidth]{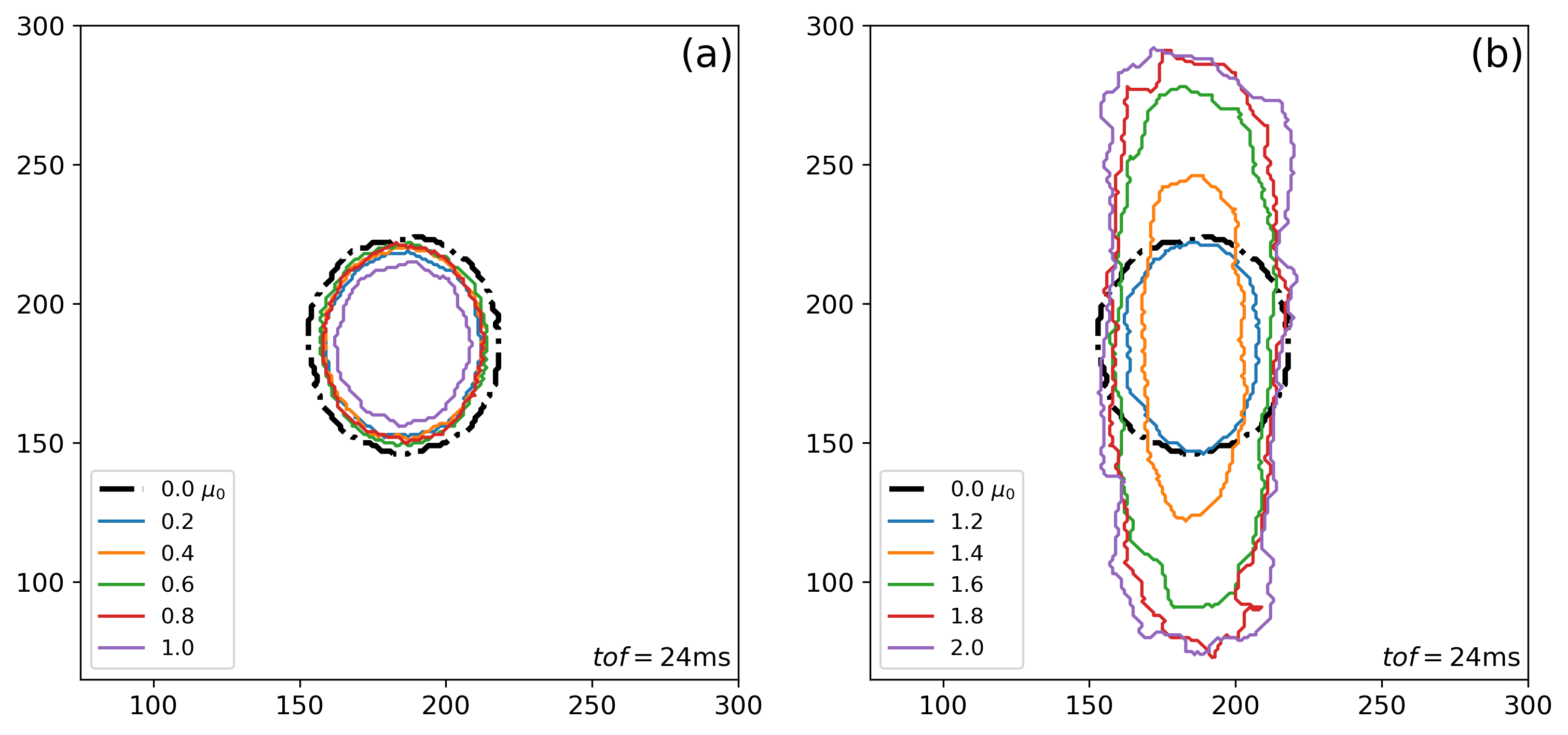}
    \caption{Perimeters determined from the binary images borders separating the regions of higher and lower optical density. They are used to determine the fractal dimension shown in Fig.\ref{fig:resF}. Note that the black dash-dot line refers to the undisturbed condensate, to help the visual inspection of the evolution as a function of the excitation amplitudes.}
    \label{fig:perims}
\end{figure}
We have used the standard algorithms and steps, well established in image processing, to analyze the general geometrical features of BEC absorption images, taken under different perturbing magnetic field conditions, to determine: its eccentricity, perimeter and the ratio between the long and short axis, without assuming any preconceived theoretical model.
\begin{figure}
    \centering
    \includegraphics[width=1.0\linewidth]{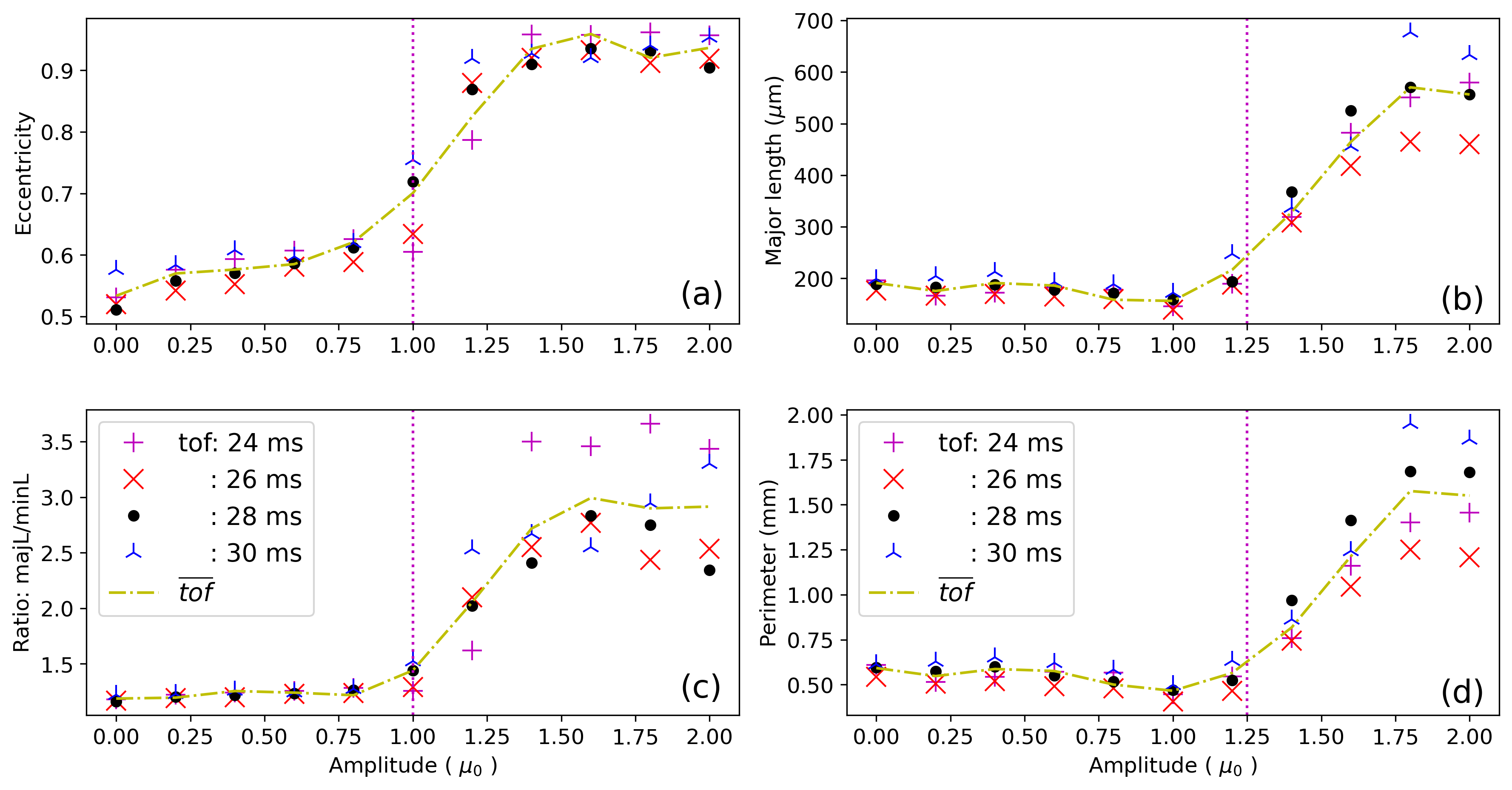}
    \caption{Initial results determined from the general analysis of the images. The orange curves work as eye guides and present the average values calculates at each particular amplitude and respective flight time.}
    \label{fig:resI}
\end{figure}
We have carried out the analysis over four different and independent datasets, taken after 24, 26, 28 and 30~ms TOF. Furthermore, we found a general trend in the evolution of the results as a function of the amplitude of the perturbing magnetic field. From 0 to 1~$\mu_0$, a very slight increase is observed, rising quickly to values a few times larger than the initial ones. The total number of atoms in the ROI region was checked and no significant fluctuations were observed.
\begin{figure}
    \centering
    \includegraphics[width=1.0\linewidth]{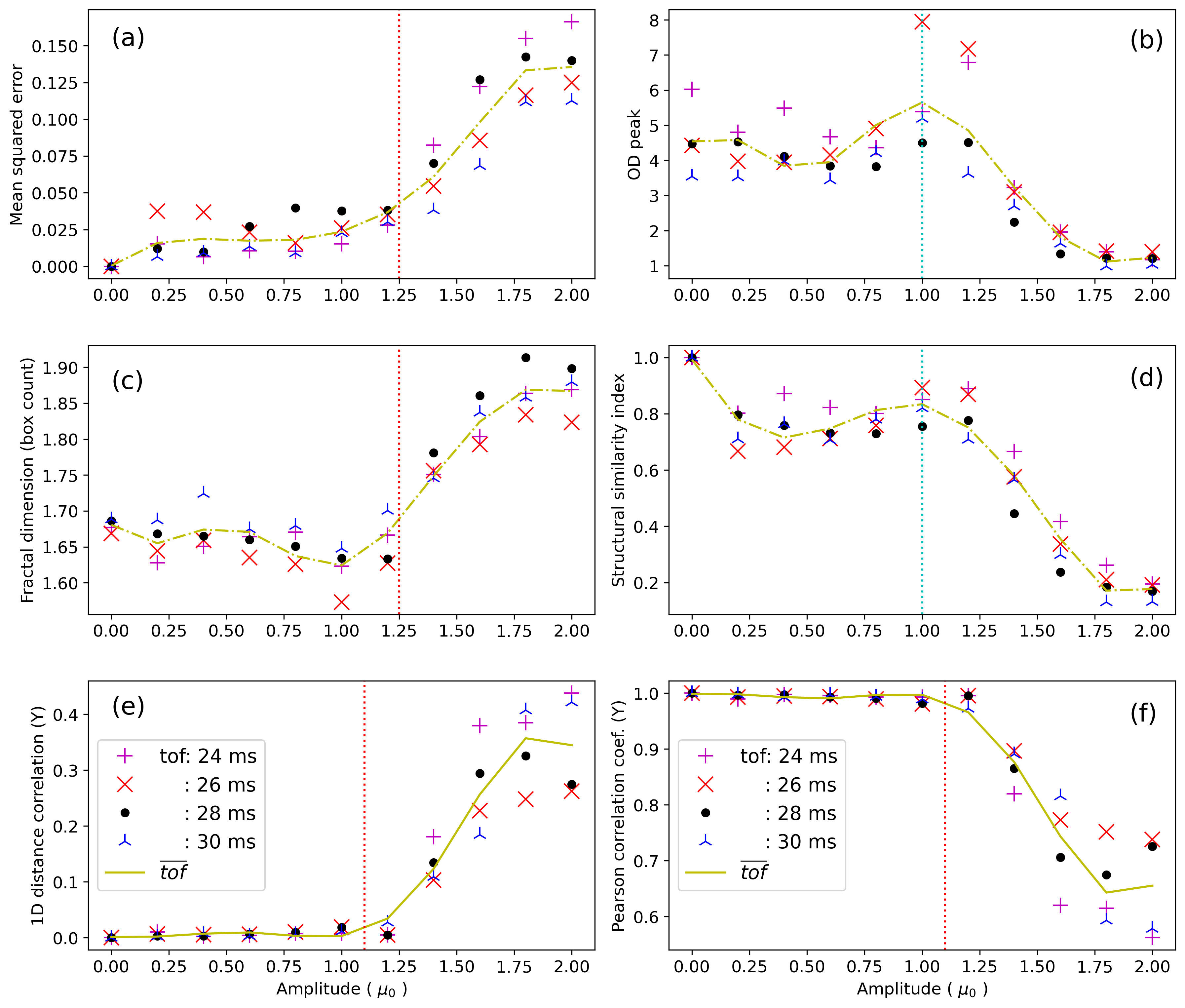}
    \caption{Main results determined from the (general) analysis of the images. The orange curves work as eye guides and present the average values calculates at each particular amplitude and respective flight time.}
    \label{fig:resF}
\end{figure}
Finally, we studied a few and more advanced methods found in the literature, which are presented in Fig.\ref{fig:perims}. Essentially, The mean-square error (MSE) evaluates the differences existing in between the pixels of two compared images~\cite{Wang09-MSE}. Whereas the structural similarity index (SSIM) searches also for similarities based on the dynamic range within pixels; i.e. it checks if the pixels in the two images line up well and have similar pixel density values~\cite{Wang04-SSIM}.

We observed a trend in the evolution of condensates as a function of increasing values of excitation amplitude. The standard behavior shows a start from a lower and relatively stable values or, sometimes, a slight growth within the range $0<A<1.0\mu_0$, and undergoing a quick increase to a new upper level. These results are consistent with what one observes in the images by inspection (Fig.\ref{fig:ImgPans}). Near $A\approx 1.0\mu_0$ we note that the central optical density reaches its top most values. And, from $A>1.2\mu_0$ onward  the central part of the cloud undergoes an fast morphing, becoming more elliptical, with the larger semi-axis (major length) growing significantly as its central density rapidly decreases. We noted that both MSE and SSIM are very consistent, though we previously expected less from the MSE measure once it tends to be less responsive to changes occurring in the gray levels. The similarity of the images also remains high until the perturbing amplitudes increase up about $1.2\mu_0$, where the values start droping sharply to lower values.

Another relevant analysis of this work is related to the fractal dimension. The profiles used to calculate it are in Fig.\ref{fig:perims}. Given the results shown in Fig.\ref{fig:resF}(c), a behavior similar to that of the previous results (Fig.\ref{fig:resI}) is observed, where the values remain relatively stable in the region of amplitudes between 0 and $1.0\mu_0$, followed by a rapid increase starting near $A\approx 1.2\mu_0$. However, this only corresponds to the final values resulting from the calculation, even though the initial intention of investigating this quantity was more ambitious. We explain: originally Lovejoy showed~\cite{Lovejoy1982}, using satellite images, that the perimeter of clouds observed in the Earth's atmosphere is a fractal, whose associated (fractal) dimension was determined $D\approx1.35$. Furthermore, in the case of classical, and fully developed turbulence, Hentschel and Procaccia showed that $D_{turb}\approx2.35$~\cite{Hentschel1983a,Hentschel1984}, suggesting that atmospheric turbulence occurring in clouds is related to surfaces of high dynamic activity that end up folding over each other in three-dimensional space, and which ultimately results in a more irregular cloud perimeters~\cite{Golitsyn22}.
Therefore, and based on visual inspection of the perimeters (borders) of the high-density regions, shown in Fig.\ref{fig:perims}, we realized that it may be plausible to use a similar methodology to investigate our images of perturbed atomic clouds. Thus, in Fig.~\ref{fig:perims}(a), we first noted a slight downsizing in the atomic cloud perimeters at smaller amplitudes, $0<A\leq 0.8\mu_0$, compared to the undisturbed condensate (black line). However, when $A=1.0\mu_0$ the perimeters become more irregular and also larger,  Fig.~\ref{fig:perims}(b). Moreover, one notices that the irregularities in the perimeters increase with the perturbing amplitudes suggesting that the system is becoming turbulent.

We also checked the correlations between the condensates existing in different degrees of excitation. We used the Pearson's correlation coefficient. As the algorithm checks the array correlation in a single dimension, we started by binning the images to acquire single dimensional profiles in the vertical and horizontal axes, which also smooths out the fast fluctuations and highlights the most striking features in the profiles. The final results are shown in the figure~\ref{fig:resF}(e-f). One may notice that the correlation variation remains similar between the different TOF values as the excitation amplitude varies. In fact, this was already expected through the direct visual inspection of the images, and noting that, as the excitation increases in the system, the atomic clouds tend to stretch along the vertical direction.
\section{Conclusions}
In conclusion, we have presented a method for using image processing to measure the shape and size of nonequilibrium Bose-Einstein condensates. We have demonstrated our strategy in a variety of experimental conditions and shown that it can be used to achieve relatively accurate measurements of the shape and size of these perturbed Bose gases. The technique presented in this article is not without its limitations. For example, it is sensitive to noise in the images and can be difficult to apply to BECs with a small number of atoms.
Nevertheless, our results provide a new way to characterize the kinematics of out-of-equilibrium BECs and may be used to study a variety of phenomena, such as the formation of vortices, the collapse of BECs, and perhaps also the dynamics of BECs existing in nonequilibrium conditions. Continued research in this area is essential for advancing our understanding of fundamental physics and developing related new technologies.
\begin{acknowledgments}
This work was funded by the São Paulo Research Foundation (FAPESP) under the grants: 2013/07276-1, 2014/50857-8, and 2022/15348-1; and also, by the National Council for Scientific and Technological Development (CNPq) under the grant 381381/2023-4.
\end{acknowledgments}
%%%%%%%%%%%
%\References
%                  
%\bibliographystyle{IEEEtran}
\bibliography{refs.bib} % Entries are in the refs.bib
%%%%%%%%%%%
\end{document}